\documentstyle[12pt,epsf]{article}

\begin{document}
\begin{titlepage}

\title{
\hfill
\parbox{5cm}{\normalsize
% ICRR-Report-???-??-??\\
\normalsize UT-850\\
}\\
\vspace{2ex}\large
The Auxiliary Mass Method beyond the Local Potential Approximation
\vspace{2ex}}
\author{\large
Kenzo Ogure\thanks{e-mail address:
 {\tt ogure@icrhp3.icrr.u-tokyo.ac.jp}}\\
 {\it Institute for Cosmic Ray Research,
   University of Tokyo}\\
   {\it  Midori-cho,  Tanashi, Tokyo 188, Japan}\\
and\\
Joe Sato\thanks{e-mail address:
 {\tt joe@hep-th.phys.s.u-tokyo.ac.jp}}\\
 {\it Department of Physics, School of Science,
   University of Tokyo}\\
   {\it   Tokyo 113-0033, Japan}}
\date{\today}

\maketitle

\begin{abstract}
    We show that the evolution equation of the effective potential in
    the auxiliary mass method corresponds to a leading approximation of a
    certain series.  This series is derived from an evolution equation
    of an effective action using a derivative expansion.  We derived
    an expression of the next-to-leading approximation of the
    evolution equation, which is a simultaneous partial differential
    equation.
\end{abstract}

\end{titlepage}

\newpage

\section{Introduction}\label{intro}
Finite temperature field theory, which is based only on the
statistical mechanism, adequately describes many physical phenomena
such as phase transitions and mass spectrum in a thermal
bath\cite{Mat,Kap,Leb}.  Its perturbation theory, however, breaks down
at a high finite temperature.  Mass squared becomes negative even in a vacuum
at the finite temperature.  This problem is solved by the ring
resummation, which adds thermal mass to zero temperature mass
beforehand\cite{Dol,Fen}.  This procedure is still insufficient yet to
make the perturbative expansion be reliable, especially around the
critical temperature.  For example, the perturbation theory indicates
that the phase transition of the $Z_2$-invariant scalar theory is of
first order incorrectly\cite{Tak}.  For the other example,
different properties are indicated by the perturbation theory and
lattice simulations in investigations of the Abelian Higgs
model\cite{Dol,Arn2,Jak,Heb,Das,Bar,Mun,Far,Kaj1,Kaj2} and the Standard
model\cite{Car,Arn,Rum,Cas,Ilg,Kar,Aok,Tet,Tet2} at large Higgs boson mass
range.

These failure of the perturbation theory is caused by bad infrared
behavior around the critical temperature\cite{Sha,Arn3}.  A loop
expansion parameter becomes $\lambda T/m$ even after the ring
resummation due to this infrared effect.  Here $\lambda$ is a small
coupling constant and $m$ is mass at the temperature $T$.  The
perturbation theory is, therefore, unreliable at a high temperature and
small mass.  This situation arises around the critical temperature of
second order phase transitions or weakly first order phase
transitions.  The auxiliary mass method controls this infrared
behavior by introducing "auxiliary
mass"\cite{DHL,IOS,OJ1,OJ2,Ogu4,Ogu3}.  We first calculate an
effective potential with the large auxiliary mass, $M\sim T$.  This
effective potential is reliable since the loop expansion parameter is
small thanks to the auxiliary mass.  We next calculate the effective
potential at true mass from this effective potential through an
evolution equation.  We solve the evolution equation of the effective
potential with respect to variation of mass squared.

We used a certain approximation to derive the evolution equation in
Ref.\cite{IOS,OJ1}.  Though we got quite good results using the
approximation, we did not have methods to improve the approximation.
In the present paper, we show that the previous evolution equation is
a leading approximation of a certain series.  We then derive a
next-to-leading evolution equation, which is a simultaneous partial
differential equation.  Though it is difficult to solve the evolution
equation at an arbitrary temperature due to numerical problems, we can,
in principle, improve the approximation systematically.

\section{Evolution Equation}

In this section, we explain our idea and derive the evolution equation
for the $Z_2$-invariant scalar theory using the auxiliary mass
method.\footnote{ Those for the other theories can be derived
similarly.} Let's consider the following Euclidean Lagrangian density
with mass squared $m^2$,
\begin{eqnarray}
    {\cal L}_{E}(\phi;m^2)=-\frac{1}{2}
     \left(\frac{\partial \phi}{\partial \tau}\right)^{2}
     -\frac{1}{2}(\mbox{\boldmath $\nabla$} \phi)^{2}
     -\frac{1}{2}m^{2}\phi^{2}
     -\frac{\lambda}{4!}(\phi^{2})^{2}
     + J_m\phi + c.t.\ .
    \label{lag}
\end{eqnarray}
Here $J_m(x)$ is an external source function which sets 
\begin{equation}
<\phi(x)>=\phi_c(x)
\label{defphic}
\end{equation}
and hence depends on $m^2$.  We set true mass squared negative,
$m^2=-\mu^2$, since we investigate the phase transition of this theory.
We assume that the coupling constant $\lambda$ is small so that the
perturbation theory be reliable at a low temperature ($T \ll m$).

Our idea is the following. The effective action for the theory,
$\Gamma [\phi_c;m^2)$, satisfies the following identity,
\begin{equation}
     \Gamma[\phi_c;-\mu^{2})=\int^{-\mu^{2}}_{M^{2}}\left(
     \frac{\partial \Gamma[\phi_c;m^2)}{\partial m^{2}} 
     \right)dm^{2}
     +\Gamma[\phi_c;M^{2}).
\label{iden}
\end{equation}
We set $M^2$ so large ($\sim T^2$) that we can calculate the initial
condition, $\Gamma[\phi_c;M^{2})$, reliably by the perturbation
theory. If we can evaluate the derivative ${\partial \Gamma(\phi_c)}/
{\partial m^{2}}$ correctly, we can calculate the effective action
accurately\footnote{ Hereafter we omit the argument $m^2$.}.

\subsection{Derivative of the effective action
with respect to mass squared
%\ $\frac{\partial \Gamma(\phi_c)}{\partial m^{2}}$ 
}
\label{sub21}
In this subsection we calculate the derivative 
$\frac{\partial \Gamma[\phi_c]}{\partial m^{2}}$.\footnote{
See the appendix in detail.}
This is formally given by\cite{IOS,OJ1},
\begin{eqnarray}
     \frac{\partial \Gamma[\phi_c]}{\partial m^{2}} 
&=& -\frac{1}{2}\int d^4x <\phi(x)^2>
\nonumber\\
&=& - \frac{1}{2}\int d^4x d^4y <\phi(x)\phi(y)>\delta(y-x)
\label{mastereq}
\\
&=&-\frac{1}{2}\int d^4x \phi_c(x)^2 - \frac{1}{2}\int d^4x d^4y \left(
\frac{\delta^2(-\Gamma[\phi_c])}{\delta\phi_c(x)\delta\phi_c(y)}
\right)^{-1}\delta(y-x),
\nonumber
\end{eqnarray}
here ``$\int d^4x$'' is an abbreviation of ``$\int^{\beta}_0 d\tau
\int_{-\infty}^{\infty} d^3x$''.  We use this notation from now on.

Since Eq.(\ref{mastereq}) is a functional equation, it can not be
solved directly.  We, therefore, limit functional space and expand
$\Gamma[\phi_c]$ in powers of
derivatives\cite{Morris,MT},\footnote{Similar derivation in ref.\cite{
Morris,MT} has a problematic point: $\exp{iqx}$ is handled as not a
distribution but an ordinary function.}
\begin{equation}
\Gamma[\phi_c] = \int d^4x \left[-V(\phi_c^2) - 
 \frac{1}{2}K_0(\phi_c^2) (\partial_0 \phi_c)^2
- \frac{1}{2}K_s(\phi_c^2) (\nabla \phi_c)^2 + \cdots\right],
\label{expansion}
\end{equation}
where $\cdots$ denotes terms with higher derivative.  Note that the
coefficient functional of $(\partial_0 \phi_c)^2$
 differs from that of $(\nabla
\phi_c)^2$ due to absence of the 4-dimensional Euclidean symmetry.  We
then expand the both side of Eq.(\ref{mastereq}) with respect to
derivative as Eq.(\ref{expansion}) and match the coefficient functionals
of each terms.  In practice, we have to truncate the series in
Eq.(\ref{expansion}) at finite terms.  In the present paper, we leave
the three terms in Eq.(\ref{expansion}). This is the next-to-leading
approximation of the derivative expansion.  We obtain
the leading approximation,
which corresponds to the evolution equation of the previous paper
Ref\cite{IOS,OJ1}, by taking $K_0 = 1$ and $\ K_s = 1$

From Eq.(\ref{expansion}), l.h.s of Eq.(\ref{mastereq}) becomes,
\begin{equation}
    - \frac{\partial \Gamma}{\partial m^{2}} = \int d^4x\left[
      \frac{\partial V}{\partial m^2} + \frac{1}{2} \frac{\partial
      K_0}{\partial m^2} (\partial_0 \phi_c)^2 +
      \frac{1}{2}\frac{\partial K_s}{\partial m^2} (\nabla \phi_c)^2
    \right] .
\label{lhs}
\end{equation}

From Eq.(\ref{expansion}), up to the second derivative
terms,\footnote{Hereafter, we also use notations like
$\overline{V}^{'}=V(\bar{\phi})$ as values of functions at constant
configuration, $\phi_c(x)=\bar\phi$.}
\begin{eqnarray}
    M_{yx}&\equiv&-\frac{\delta^2\Gamma[\phi_c]}
    {\delta\phi_c(x)\delta\phi_c(y)}
    \nonumber\\ 
    &=&\delta(y-x)\left[V^{''}(\phi_c(y))\right.
    \nonumber\\ 
    &&-\left\{\frac{1}{2}K_0^{''}(\phi_c(y))(\partial_0\phi_c(y))^2
      +K_0^{'}(\phi_c(y))\partial_0\phi_c(y) \partial_{y_0}\right.
    \nonumber\\ 
    &&\left.+K_0^{'}(\phi_c(y))
      \partial_0^2\phi_c(y)+K_0(\phi_c(y))\partial_{y_0}^2\right\}
\label{true_propagator1}
\\ &&\left.  -\left\{K_0\leftrightarrow K_s,
    \partial_{y_0}\leftrightarrow \nabla_y\right\} \right] .
\nonumber\\ 
&=& \delta(y-x)(-\overline{K}_0 \partial_{y_0}^2 -
\overline{K}_s \nabla_y^2 +\overline{V}'') \nonumber\\ 
&&+\delta(y-x)
\left[{\widetilde{V}^{''}(\phi_c(y))} \right.\nonumber\\ 
&&-\left\{\frac{1}{2}K_0^{''}(\phi_c(y)) (\partial_0\phi_c(y))^2
  +K_0^{'}(\phi_c(y)) \partial_0\phi_c(y)\partial_{y_0}
\right.\nonumber\\ 
&&+\left.K_0^{'}(\phi_c(y))
  \partial_0^2\phi_c(y)+\widetilde
  K_0(\phi_c(y))\partial_{y_0}^2\right\}
\label{true_propagator2}
\\
&&-\left.
\left\{K_0\leftrightarrow K_s, \partial_{y_0}\leftrightarrow \nabla_y\right\}
\right] 
\nonumber\\
&=& A_{yx}-B_{yx},
\label{propagator_division}\\
A_{yx}&\equiv&\delta(y-x)(-\overline{K}_0 \partial_{y_0}^2 -
\overline{K}_s \nabla_y^2 +\overline{V}'')
\nonumber\\
B_{yx}&\equiv& A_{yx}-M_{yx}
\nonumber\\
\widetilde{V}^{''}(\phi_c(y))
&\equiv&V^{''}(\phi_c(y))-\overline V^{''}
\nonumber\\
\widetilde K_0(\phi_c(y))&\equiv&K_0(\phi_c(y))-\overline K_0
\nonumber
\end{eqnarray}
Here, we divide $M_{yx}$ into two parts, $A_{yx}$ and $B_{yx}$, which
remains finite and vanishes at $\phi_c(x)=\bar\phi$ respectively.
Details of the following calculation are explained in the appendix.  
The right hand side of Eq.(\ref{mastereq}) can now be expanded around
$\phi_c(x)=\bar\phi$ up to the second derivative as follows,
\begin{eqnarray}
&&\int d^4x d^4y 
(\frac{\delta^2(-\Gamma[\phi_c])}{\delta\phi_c(x)\delta\phi_c(y)})^{-1}
\delta(y-x)
\nonumber\\
&=&\int d^4x (A^{-1})_{xx}
\nonumber\\
&&+\int d^4x d^4y d^4z (A^{-1})_{xy} B_{yz}
(A^{-1})_{zx}
\nonumber\\
&&+\int d^4x d^4y d^4z d^4u d^4v (A^{-1})_{xy} B_{yz}
(A^{-1})_{zu}B_{uv} (A^{-1})_{vx}
\nonumber\\
&& + \cdots
\nonumber\\
&=&\int_{px}\frac{1}{\nu}
\nonumber\\
&&+\int_{px}\frac{1}{\nu^2}
\left[-\frac{1}{2}\left\{\overline K_0''(\partial_0\phi_c)^2+
\overline K_s''(\nabla\phi_c)^2\right\}\right]
\nonumber\\
&&+\int_{px}\frac{1}{\nu^3}
\left[\left\{\overline K_0'p_0(\partial_0\phi_c)
+\overline K_s'p_i(\partial_i\phi_c)\right\}^2\right.
\nonumber\\
&&\ \ \ \ \ \ \ \ \ 
\left.+\ 2\left\{\overline K'_0 (\partial_0 \phi_c)^2
    +\overline K'_s(\nabla\phi_c)^2\right\}\nu'
\right]
\label{rhs}
\\
&&\int_{px}\frac{1}{\nu^4}\left[-2\left\{\overline K_0'p_0\partial_0\phi_c
+\overline K_s'p_i\partial_i\phi_c\right\}
\left\{\overline K_0p_0\partial_0\phi_c
+\overline K_s p_i\partial_i\phi_c\right\}\nu'
\right.
\nonumber\\
&&\left.\ \ \ \ \ \ \ 
-\frac{1}{2}
\left\{\nu'\right\}^2
\left\{\overline K_0(\partial_0\phi_c)^2+
\overline K_s(\nabla\phi_c)^2\right\}
\right]
\nonumber\\
&&+\cdots +(terms\ which\ vanish\ at\ \phi_c(x)=\bar\phi\ like\
\widetilde V\times (something)),\nonumber
\end{eqnarray}
where $\nu=\overline K_0 p_0^2+ \overline K_s {\bf p}^2 +\overline
V''$, $\nu'=\overline K'_0 p_0^2+ \overline K'_s {\bf p}^2 +\overline
V'''$ (with $p_0=2\pi n T (n=0,\pm 1,\cdots)$ and ${\bf p}^2 \equiv
\sum_{i=1}^3 p_i^2$) and ``$\int_{px}$'' is an abbreviation of
``$T\sum_{n=-\infty}^\infty \int d^3p/(2\pi)^3\int d^4x$''.

We match the both side of Eq.(\ref{mastereq}) using Eq.(\ref{lhs}) and
Eq.(\ref{rhs}) and equate the coefficient functionals of
$(\partial\phi_c)$ respectively.  After the matching, we put $\phi_c
(x)=\bar\phi$ and get the following simultaneous partial differential
equation,
\begin{eqnarray}
    \frac{\partial \overline V}{\partial m^2}&=&\frac{1}{2}
    \bar\phi^2+\frac{1}{2}\int_p \frac{1}{\nu}
\label{eveqV}\\
\frac{\partial \overline K_0}{\partial m^2}&=&-\frac{1}{2}
\overline K_0''\int_p\frac{1}{\nu^2}
\nonumber\\
&&+ \overline K_0^2 \int_p \frac{p_0^2}{\nu^3}+2 \overline K_0'
\int_p \frac{\nu'}{\nu^3}
\label{eveqK0}\\
&&-2 \overline K_0'\overline K_0\int_p p_0^2 \frac{\nu'}{\nu^4}
-\frac{1}{2} \overline K_0\int_p\frac{\{\nu'\}^2}{\nu^4}
\nonumber\\
\frac{\partial \overline K_s}{\partial m^2}&=&-\frac{1}{2}
\overline K_s''\int_p\frac{1}{\nu^2}
\nonumber\\
&&+\frac{ \overline K_s^2}{3} \int_p \frac{{\bf p}^2}
{\nu^3}+2 \overline K_s'\int_p \frac{\nu'}{\nu^3}
\label{eveqKs}\\
&&-\frac{2}{3} \overline K_s'\overline K_s\int_p {\bf p}^2 \frac{\nu'}{\nu^4}
-\frac{1}{2}\overline K_s\int_p\frac{\{\nu'\}^2}{\nu^4},
\nonumber
\end{eqnarray}
where, ``$\int_p$'' is an abbreviation of ``$T\sum_{n=-\infty}^{\infty}
\int d^3p/(2\pi)^3$''.  Since we put $\phi_c (x)=\bar\phi$ finally, we
do not have contribution from the terms in Eq.(\ref{rhs}) which vanish
at $\phi_c(x)=\bar\phi$. This is the evolution equation of the
next-to-leading approximation of the derivative expansion.

\subsection{Initial Condition}\label{sub22}
We can calculate the initial condition $\Gamma[\bar\phi;M^2)$ using
the perturbation theory within one-loop level thanks to the large
auxiliary mass, $M\sim T$.  The effective potential, $V(\bar\phi;M^2)$,
is calculated within the one-loop approximation as follows,
\begin{equation}
V(\bar\phi;M^2)=\frac{1}{2} M^2 \bar\phi + \frac{\lambda}{24} \bar\phi^4
+\frac{1}{2}\int_p\log(p_0^2+{\bf p}^2 + M^2 + \frac{\lambda}{2}\bar\phi^2).
\label{Vini}
\end{equation}

\begin{figure}
\unitlength=1cm
\begin{picture}(8,4)
\unitlength=1mm
\centerline{
\epsfxsize=10cm
\epsfbox{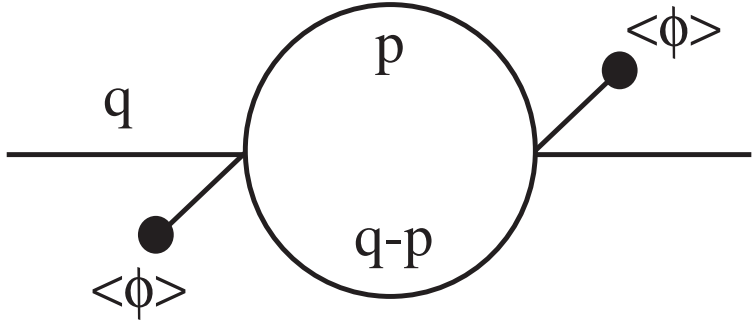} 
}
\end{picture}
\caption{Diagram of the external-momentum dependent self-energy at one-loop.}
\label{kfig}
\end{figure}

The loop correction to $K_0(\bar\phi;M^2)$ and $K_s(\bar\phi;M^2)$
comes from the self-energy graph in Fig.\ref{kfig} at one-loop level,
which depends on external momentum,
\begin{equation}
    \Pi(q_0^2,{\bf q}^2)= \frac{\lambda^2 \bar\phi^2}{2}
    \int_p\frac{1}{p_0^2+{\bf p}^2+M^2+\frac{\lambda\bar\phi^2}{2}}
    \frac{1}{(q_0-p_0)^2+({\bf
    q-p})^2+M^2+\frac{\lambda\bar\phi^2}{2}}.
\end{equation}
The initial conditions of $K_0(\bar\phi;M^2)$ and $K_s(\bar\phi;M^2)$
are given by the coefficients of $-q_0^2 $ and $-{\bf q}^2$
respectively. We, therefore, get\footnote{Strictly speaking, we first
put $q_i=0$ then put $q_0=0$ in taking the limit ${q_i=0}$ and
${q_0=0}$.  For the function $K_0$ to be defined this order is
essential, while for $K_s$ the order of taking the limit gives no
difference.  See ref. \cite{QGP}.}
\begin{eqnarray}
K_0(\bar\phi;M^2)&=&1-
\left.\frac{d\Pi(q_0^2,{\bf q}^2)}{dq_0^2}\right|_{q_i=0,q_0=0}
\label{K0ini}\\
&=& 1-\frac{\lambda^2\bar\phi^2}{2} \int_p 
\left[\frac{1}{(p_0^2+{\bf p}^2+M^2+
\frac{\lambda\bar\phi^2}{2})^3}-\frac{M^2+
\frac{\lambda\bar\phi^2}{2}}{(p_0^2+{\bf p}^2+M^2+
\frac{\lambda\bar\phi^2}{2})^4}\right],
\nonumber\\
K_s(\bar\phi;M^2)&=&1-\left.\frac{d\Pi(q_0^2,{\bf q}^2)}
{d{\bf q}^2}\right|_{q_0=0,q_i=0}
\label{Ksini}\\
&=& 1+\frac{\lambda\bar\phi^2}{6} \int_p 
\frac{1}{(p_0^2+{\bf p}^2+M^2+
\frac{\lambda^2\bar\phi^2}{2})^3}.
\nonumber
\end{eqnarray}

We have calculated the evolution equation in
\S\ref{sub21} and the initial condition in \S\ref{sub22}. Since some
of them, e.g.. the one-loop effective potential, has an ultraviolet
divergence, we have to renormalize it by the counter terms. Instead of
considering this contribution, we simply assume that the
renormalization effect is small and discard it.  The contribution is,
actually, small comparing with the finite temperature contribution
around a critical temperature in most cases.  We thus need to deal with
only the temperature dependent pieces in the integrals of
$\frac{\partial\Gamma}{\partial m^2}$ and the initial condition.  To
do so, we only perform the following replacement in
Eq.(\ref{eveqV}),(\ref{eveqK0}),(\ref{eveqKs}),(\ref{Vini}),(\ref{K0ini})
,(\ref{Ksini}),
\begin{eqnarray}
\int_p\ \ \ \ \ 
\hbox{with}\ \ \ \ \ 
2 \int \frac{d^4p}{(2\pi)^4 i}\frac{1}{\exp (p_0/T) - 1}.
\nonumber
\end{eqnarray}

We note that under the local potential approximation that $K_0=K_s=1$,
the evolution equation of the effective potential reproduce our
previous equation.\cite{IOS,OJ1}.

\section{Results and Summary}
\label{sec:Results and Summary}
We solved the simultaneous evolution equation Eq.(\ref{eveqV}),
Eq.(\ref{eveqK0}) and Eq.(\ref{eveqKs}) with the initial condition
Eq.(\ref{Vini}), Eq.(\ref{K0ini}) and Eq.(\ref{Ksini}) numerically.
We used an extended Crank-Nicholson method, which is explained in the 
appendix of Ref\cite{OJ1}, to solve the partial differential
equation.  We could solve the equation at most temperatures above the
critical one.  Unfortunately, we can not, however, solve the equation
at temperatures very close to the critical one due to numerical
problems.  We tried several improvements of the numerical method but
failed.  We, therefore, show only obtained results in the present
paper. This equation may be solved at arbitrary temperature if
excellent numerical methods to solve a partial differential equation
are invented. Or ability
of computers progresses highly so that we can calculate with much higher 
precision.  We use a mass unit, $\mu =1$

\begin{figure}
\unitlength=1cm
\begin{picture}(8,4)
\unitlength=1mm
\centerline{
\epsfxsize=10cm
\epsfbox{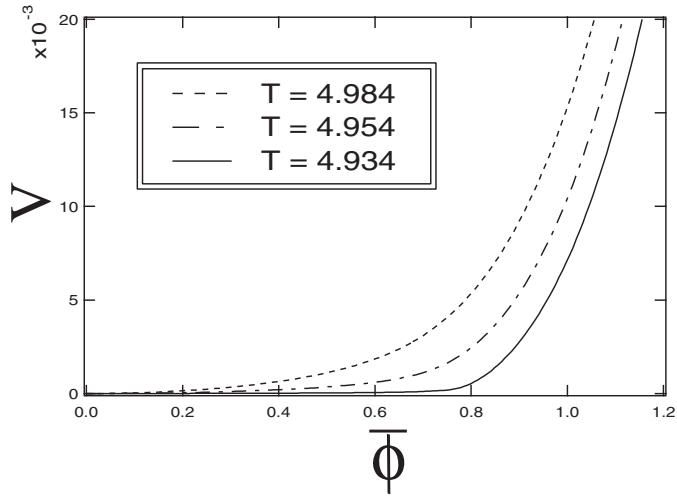} 
}
\end{picture}
\caption{Effective potential around the critical temperature ($\lambda=1$).}
\label{V}
\end{figure}

We show the effective potential above the critical temperature in
Fig.\ref{V} for $\lambda =1$.  We observe behavior of a second
order phase transition up to this temperature.  This effective potential
seems to be cone-shaped.  The critical temperature is estimated to be
lower than that of the local potential
approximation by $2\%$.  

\begin{figure}
    \unitlength=1cm
\begin{picture}(8,8)
\unitlength=1mm
\centerline{
\epsfxsize=10cm
\epsfbox{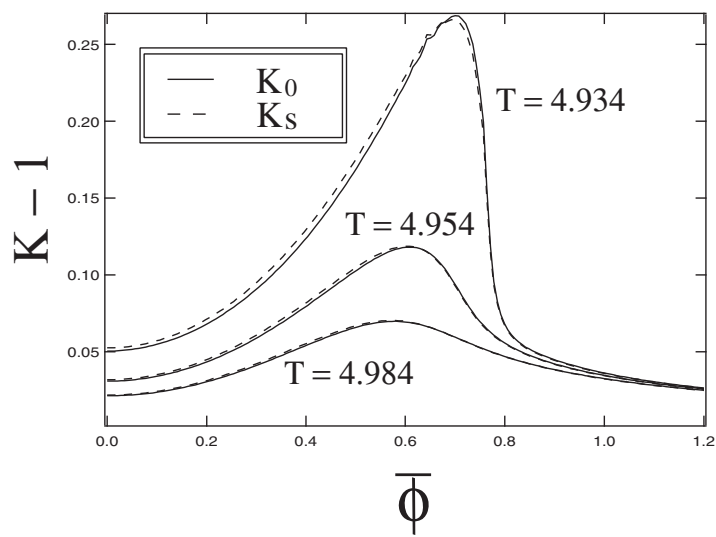} 
}
\end{picture}
\caption{The coefficient functions of the second derivative terms in
the effective action, $K_0$ and $K_s$ ($\lambda=1$).}
\label{K}
\end{figure}

We show $K_0$ and $K_s$ in Fig.\ref{K} for $\lambda =1$. They are very
similar in spite of the violation of the Lorenz invariance in this
case while the initial condition is quite different.

In summary, we derived an evolution equation of the effective action
with respect to the mass squared in the $Z_2$-invariant scalar theory.
We then approximated the effective action by the derivative expansion.
We showed that the previous evolution equation of the effective potential
can be derived as the leading approximation, the local potential
approximation.  We next derived the evolution equation of the
next-to-leading approximation which is the simultaneous partial
differential equation.  We finally solved the equation numerically.
Though we could solve it at most temperatures above the critical one,
we could not do it under a certain temperature very close to the
critical one unfortunately.  However, this equation may be solved at
arbitrary temperatures if excellent numerical methods to solve a partial
differential equation are invented or ability of computers progresses
highly.  Anyway, we constructed the systematic method to improve the
auxiliary mass method in principle.

We are suported by JSPS.

\newpage

\begin{center}
{\Large Appendix}
\end{center}
\appendix
In this appendix,     we show the derivation of ${\partial\Gamma}/{\partial
m^2}$ in detail. Some notations are given in the chapter 2.  The
effective action $\Gamma[\phi_c]$ for the Lagrangian Eq.(\ref{lag}) is 
defined as usual,
\begin{eqnarray}
\Gamma[\phi_c]&\equiv&
W[J_m]-\int d^4x J_m(x) \phi_c(x),
\label{defgamma}\\
W[J_m]&\equiv&\log (Z[J_m]),
\nonumber\\
Z[J_m]&\equiv&\int
{\cal D}[\phi]\exp\left(\int d^4x {\cal L}_E\right).
\nonumber
\end{eqnarray}
The equation(\ref{mastereq}) is, then, derived as follows,
\begin{eqnarray}
    \frac{\partial \Gamma[\phi_c]}{\partial m^{2}} &=&
\frac{1}{Z[J_m]} \int {\cal D}[\phi] \left[\int \left\{d^4x 
\frac{1}{2}(-\phi(x)^2)
+\frac{\partial J_m(x)}{\partial m^2}\phi(x)\right\}
\exp\left\{\int d^4x {\cal L}_E\right\} \right]
\nonumber\\
&&-\int d^4x \frac{\partial J_m(x)}{\partial m^2} \phi_c(x)
\nonumber\\
&=&
-\frac{1}{2}\int d^4x <\phi(x)^2>
\nonumber\\
&=&
-\frac{1}{2}\int d^4x\ \phi_c(x)^2
-\frac{1}{2}\int d^4x (M)_{xx}^{-1}.
\label{M_inverse_trace}
\end{eqnarray}

Since Eq.(\ref{mastereq}) is a functional equation, it can not be
solved directly.  We, therefore, limit functional space and expand
$\Gamma[\phi_c]$ in powers of
derivatives\cite{Morris,MT}, (see Eq.(\ref{expansion}))
\begin{eqnarray}
    \Gamma[\phi_c] = \int d^4x \left[-V(\phi_c^2) -
      \frac{1}{2}K_0(\phi_c^2) (\partial_0 \phi_c)^2 -
      \frac{1}{2}K_s(\phi_c^2) (\nabla \phi_c)^2 + \cdots\right],
    \nonumber
\end{eqnarray}
where ``$\cdots$'' denotes terms with higher derivative which are
omitted here and equating the coefficient functionals
of $(\partial\phi_c)$ in
the both side of Eq.(\ref{mastereq}).

The left hand side of Eq.(\ref{M_inverse_trace}) is calculated to be
Eq.(\ref{lhs}) by simply differentiating $\Gamma$ with respect to
$m^2$,
\begin{eqnarray}
    - \frac{\partial \Gamma}{\partial m^{2}} =
\int d^4x\left[ \frac{\partial V}{\partial m^2} + \frac{1}{2}
\frac{\partial K_0}{\partial m^2} (\partial_0 \phi_c)^2
+ \frac{1}{2}\frac{\partial K_s}{\partial m^2} (\nabla \phi_c)^2 
\right] .
\nonumber
\end{eqnarray}

It is very complicated to calculate r.h.s. of
Eq.(\ref{M_inverse_trace}). First we calculate $M_{yx}$. The
derivative of the effective action $\Gamma[\phi_c]$ with respect
to $\phi_c(x)$ is,
\begin{eqnarray}
\frac{\delta\Gamma[\phi_c]}{\delta\phi_c(x)}=&
-&V^{'}(\phi_c(x))\\
&&+\frac{1}{2}K_0^{'}(\phi_c(x))
(\partial_0\phi_c(x))^2+K_0(\phi_c(x))(\partial_0^2\phi_c(x))
\nonumber\\
&&+\left\{K_0\leftrightarrow K_s, \partial_0\leftrightarrow \nabla\right\}.
\nonumber
\end{eqnarray}
Here, we assume that we can make a partial integral
freely without a surface term.

We define the operator $M_{xy}$ in the following sense. An operator
$O_{xy}$ defined through functional derivative with
respect to $\phi(x)$, say $\delta F(\phi(y))/\delta\phi(x)$, 
acts on any appropriate test function, say
$T(y)$,
\begin{eqnarray}
OT(x)&\equiv& \int d^4y 
\left\{\frac{\delta}{\delta\phi(x)}F(\phi(y))\right\}
T(y).\nonumber
\end{eqnarray}
In particular, if $F(\phi(y))$ contains the derivative of $\phi(y)$,
say $F(\phi(y))=G(\phi(y))\partial\phi(y)$,
\begin{eqnarray}
OT(x)&\equiv& \int d^4y 
\left\{\frac{\delta}{\delta\phi(x)}G(\phi(y))\partial\phi(y)\right\}
T(y).\nonumber\\
&\equiv&
\int d^4y \delta(y-x)
\left[G^{'}(\phi(y))\partial\phi(y)T(y)
-\partial \left\{ G(\phi(y))T(y)\right\}\right].
\end{eqnarray}

In order to obtain Eq.(\ref{true_propagator1}) and to determine
the inverse of $M_{yx}$ around $\bar\phi$, we divide the
Eq.(\ref{true_propagator1}) into two pieces
(Eq.(\ref{propagator_division})),
\begin{eqnarray}
M_{yx}&=& A_{yx}-B_{yx},\nonumber \\
A_{yx}&=&\delta(y-x)(-\overline{K}_0 \partial_0^2 
-\overline{K}_s \nabla^2 +\overline{V}''),
\nonumber\\
B_{yx}&=&-\delta(y-x)
\left[\widetilde{V}^{''}(\phi_c(y))
\right.\nonumber\\
&-&\left\{\frac{1}{2}K_0^{''}(\phi_c(y))
(\partial_0\phi_c(y))^2
+K_0^{'}(\phi_c(y))
\partial_0\phi_c(y)\partial_{y_0}
\right.\nonumber\\
&+&\left.K_0^{'}(\phi_c(y))
\partial_0^2\phi_c(y)+\widetilde K_0(\phi_c(y))\partial_{y_0}^2\right\}
\nonumber
\\
&&-\left.
\left\{K_0\leftrightarrow K_s, \partial_{y_0}\leftrightarrow \nabla_y\right\}
\right] .
\nonumber
\end{eqnarray}
The inverse of $M_{yx}$ is, then, expanded as,
\begin{eqnarray}
(M^{-1})_{xy}&=&\{(A^{-1})(\sum_{n=0}^{n=\infty} (BA^{-1})^n)\}_{xy}.
\label{M_inverse}
\end{eqnarray}
Here, the multiplication of the "matrix" $A$ and $B$ is taken in the 
following sense,
\begin{eqnarray}
(AB)_{xy} = \int d^4z A_{xz} B_{zy}.
\nonumber
\end{eqnarray}
The inverse of $A_{xy}$ is calculated easily,
\begin{eqnarray}
(A^{-1})_{xy} = \int_p \frac{1}{\nu_p}\exp\left\{ip(x-y)\right\},
\label{A_inverse}
\end{eqnarray}
where  $\nu_p=\overline{K}_0 p_0^2+ \overline{K}_s {\bf p}^2 
+\overline{V}''$.

We need terms with $(\partial\phi_c)^n (n=0,2)$
to evaluate r.h.s of Eq.(\ref{M_inverse_trace}).
Such terms are contained only in the first
three terms in Eq.(\ref{M_inverse}),
\begin{eqnarray}
(M^{-1})_{xy}
&=&(A^{-1})_{xy}
\nonumber\\
&&+ (A^{-1}BA^{-1})_{xy}
\label{M_inverse2}\\
&&+(A^{-1} B A^{-1} B A^{-1})_{xy}.
\nonumber
\end{eqnarray}
The relevant terms in Eq.(\ref{M_inverse_trace}) are, therefore, the
following,
\begin{eqnarray}
\int d^4x (M^{-1})_{xx}
&=&\int d^4x (A^{-1})_{xx}
\label{first}\\
&&+\int d^4x  (A^{-1}BA^{-1})_{xx}
\label{second}\\
&&+\int d^4x (A^{-1} B A^{-1} B A^{-1})_{xx}.
\label{third}
\end{eqnarray}
We next calculate Eq.(\ref{first})-(\ref{third})
up to terms with the second derivative of $\phi_c(x)$.
From Eq.(\ref{A_inverse}) the first term (\ref{first})
is easy to calculate,
\begin{eqnarray}
\int d^4x (A^{-1})_{xx}=\int d^4x \int_p \frac{1}{\nu_p}.
\label{Axx}
\end{eqnarray}
The second term Eq.(\ref{second}) is
\begin{eqnarray}
&&\int d^4x  (A^{-1}BA^{-1})_{xx}
\nonumber\\
&=&\int d^4x \int d^4y \int d^4z
\int_p \frac{1}{\nu_p}\exp\left\{ip(x-y)\right\}
\nonumber\\
&&\delta(y-z)
\left[-\widetilde{V}^{''}(\phi_c(y))
+\left\{\frac{1}{2}K_0^{''}(\phi_c(y))
(\partial_0\phi_c(y))^2
+K_0^{'}(\phi_c(y))
\partial_0\phi_c(y)\partial_{y_0}
\right.\right.\nonumber\\
&&\left.+\left.K_0^{'}(\phi_c(y))
\partial_0^2\phi_c(y)+\tilde K_0(\phi_c(y))\partial_{y_0}^2\right\}
+\left\{K_0\leftrightarrow K_s, \partial_{y_0}\leftrightarrow \nabla_y\right\}
\right] 
\nonumber\\
&&\int_q \frac{1}{\nu_q}\exp\left\{iq(z-x)\right\}
\nonumber\\
&&\hbox{(integrating over $z$.)}
\nonumber\\
&=&\int d^4x \int d^4y 
\int_p \int_q \frac{1}{\nu_p} \frac{1}{\nu_q}
\exp\left\{ip(x-y)\right\} \exp\left\{iq(y-x)\right\}
\nonumber\\
&&\left[-\widetilde{V}^{''}(\phi_c(y))
+\left\{\frac{1}{2}K_0^{''}(\phi_c(y))
(\partial_0\phi_c(y))^2
+K_0^{'}(\phi_c(y))
\partial_0\phi_c(y)iq_0
\right.\right.\nonumber\\
&&\left.+\left.K_0^{'}(\phi_c(y))
\partial_0^2\phi_c(y)-\tilde K_0(\phi_c(y))q_0^2\right\}
+\left\{K_0\leftrightarrow K_s, \partial_0\leftrightarrow \nabla,
q_0 \leftrightarrow {\bf q}\right\}
\right]
\nonumber\\
&&\hbox{(integrating over $x$ and $q$ then replacing $y$ by $x$.)}
\nonumber\\
&=&\int d^4x \int_p \frac{1}{\nu_p^2} 
\nonumber\\
&&\left[-\widetilde{V}^{''}(\phi_c(x))
+\left\{\frac{1}{2}K^{''}_0 (\phi_c(x))
(\partial_0\phi(x))^2
+K_0^{'}(\phi_c(x))
\partial_0\phi(x)ip_0
\right.\right.\nonumber\\
&&\left.+\left.K_0^{'}(\phi_c(x))
\partial_0^2\phi(x)-\tilde K_0(\phi_c(x))p_0^2\right\}
+\left\{K_0\leftrightarrow K_s, \partial_0\leftrightarrow \nabla,
p_0 \leftrightarrow {\bf p}\right\}
\right]
\nonumber\\
&&\hbox{(partially integrating the fourth term.)}
\nonumber\\
&=&\int d^4x \int_p \frac{1}{\nu_p^2} 
\left[-\widetilde{V}^{''}(\phi_c(x))
-\left\{\frac{1}{2}K^{''}_0 (\phi_c(x))
(\partial_0\phi_c(x))^2+\tilde K_0(\phi_c(x))p_0^2\right\}
\right.
\nonumber\\
&&\left.
-\left\{K_0\leftrightarrow K_s, \partial_0\leftrightarrow \nabla,
p_0 \leftrightarrow {\bf p}\right\}
\right].
\nonumber\\
&&
\nonumber\\
&=&\int d^4x \int_p \frac{1}{\nu_p^2} 
\left[-\left\{\frac{1}{2} \overline{K}_0''(\partial_0\phi_c)^2
+\frac{1}{2} \overline{K}_s''(\nabla\phi_c)^2
\right\}\right]\nonumber \\
&&+(terms\ which\ vanish\ at\ \phi_c(x)=\bar\phi\ like\
\widetilde V''\times (something) ).
\label{ABAxx}
\end{eqnarray}
Since we set $\phi_c(x)=\bar\phi$ after the matching of
Eq.(\ref{mastereq}), we have no
contribution to the evolution equation from the terms which vanish at
$\phi_c(x)=\bar\phi$. We thus leave only terms which remain finite at
$\phi_c(x)=\bar\phi$ from now on. The third term (\ref{third}) is very
complicated to evaluate,
\begin{eqnarray}
&&\int d^4x (A^{-1} B A^{-1} B A^{-1})_{xx}
\nonumber\\
&=&\int d^4x \int d^4y \int d^4z \int d^4u \int d^4v
\int_p \frac{1}{\nu_p}\exp\left\{ip(x-y)\right\}
\nonumber\\
&&\delta(y-z)
\left[-\widetilde{V}^{''}(\phi_c(y))
+\left\{\frac{1}{2}K_0^{''}(\phi_c(y))
(\partial_0\phi_c(y))^2
+K_0^{'}(\phi_c(y))
\partial_0\phi_c(y)\partial_{y_0}
\right.\right.\nonumber\\
&&\left.+\left.K_0^{'}(\phi_c(y))
\partial_0^2\phi_c(y)+\tilde K_0(\phi_c(y))\partial_{y_0}^2\right\}
+\left\{K_0\leftrightarrow K_s, \partial_{y_0}\leftrightarrow \nabla_y\right\}
\right] 
\nonumber\\
&&\int_q \frac{1}{\nu_q}\exp\left\{iq(z-u)\right\}
\nonumber\\
&&\delta(u-v)
\left[-\widetilde{V}^{''}(\phi_c(u))
+\left\{\frac{1}{2}K_0^{''}(\phi_c(u))
(\partial_0\phi_c(u))^2
+K_0^{'}(\phi_c(u))
\partial_0\phi_c(u)\partial_{u_0}
\right.\right.\nonumber\\
&&\left.+\left.K_0^{'}(\phi_c(u))
\partial_0^2\phi_c(u)+\tilde K_0(\phi_c(u))\partial_{u_0}^2\right\}
+\left\{K_0\leftrightarrow K_s, \partial_{u_0}\leftrightarrow \nabla_u\right\}
\right] 
\nonumber\\
&&\int_r \frac{1}{\nu_r}\exp\left\{ir(v-x)\right\}
\nonumber\\
&&\hbox{(integrating over $x,z,v,r$ and replacing $u$ with $x$)}
\nonumber\\
&=&\int d^4x \int d^4y \int_p \int_q \frac{1}{\nu_p^2}\frac{1}{\nu_q} 
\exp\left\{i(p-q)(x-y)\right\}
\nonumber\\
&&\left[-\widetilde{V}^{''}(\phi_c(x))
+\left\{\frac{1}{2}K^{''}_0 (\phi_c(x))
(\partial_0\phi_c(x))^2
+K_0^{'}(\phi_c(x))
\partial_0\phi_c(x)ip_0
\right.\right.\nonumber\\
&&\left.+\left.K_0^{'}(\phi_c(x))
\partial_0^2\phi_c(x)-\tilde K_0(\phi_c(x))p_0^2\right\}
+\left\{K_0\leftrightarrow K_s, \partial_0\leftrightarrow \nabla,
p_0 \leftrightarrow {\bf p}\right\}
\right]
\nonumber\\
&\times&\left[-\widetilde{V}^{''}(\phi_c(y))
+\left\{\frac{1}{2}K_0^{''}(\phi_c(y))
(\partial_0\phi_c(y))^2
+K_0^{'}(\phi_c(y))
\partial_0\phi_c(y)iq_0
\right.\right.\nonumber\\
&&\left.+\left.K_0^{'}(\phi_c(y))
\partial_0^2\phi_c(y)-\tilde K_0(\phi_c(y))q_0^2\right\}
+\left\{K_0\leftrightarrow K_s, \partial_0\leftrightarrow \nabla,
q_0 \leftrightarrow {\bf q}\right\}
\right].
\nonumber\\
&&\hbox{(partially integrating the term with $\partial_0^2\phi$.)}
\nonumber\\
&=&\int d^4x \int d^4y \int_p \int_q \frac{1}{\nu_p^2}\frac{1}{\nu_q} 
\exp\left\{i(p-q)(x-y)\right\}
\nonumber\\
&&\left[-\widetilde{V}^{''}(\phi_c(x))
+\left\{-\frac{1}{2}K^{''}_0 (\phi_c(x))
(\partial_0\phi_c(x))^2
+K_0^{'}(\phi_c(x))
\partial_0\phi_c(x)iq_0
\right.\right.\nonumber\\
&&\left.-\left.\tilde K_0(\phi_c(x))p_0^2\right\}
+\left\{K_0\leftrightarrow K_s, \partial_0\leftrightarrow \nabla,
p_0 \leftrightarrow {\bf p}, q_0 \leftrightarrow {\bf q}\right\}
\right]
\nonumber\\
&\times&\left[-\widetilde{V}^{''}(\phi_c(y))
+\left\{-\frac{1}{2}K_0^{''}(\phi_c(y))
(\partial_0\phi_c(y))^2
+K_0^{'}(\phi_c(y))
\partial_0\phi_c(y)ip_0
\right.\right.\nonumber\\
&&\left.-\left.\tilde K_0(\phi_c(y))q_0^2\right\}
+\left\{K_0\leftrightarrow K_s, \partial_0\leftrightarrow \nabla,
p_0 \leftrightarrow {\bf p}, q_0 \leftrightarrow {\bf q}\right\}
\right].
\nonumber
\end{eqnarray}

By the following variable exchange,
\begin{eqnarray}
\left\{
\begin{array}{lll}
p &\longrightarrow& p+q\\
q &\longrightarrow& p
\end{array}
\right.,
\nonumber
\end{eqnarray}
we obtain,
\begin{eqnarray}
&=&\int d^4x \int d^4y \int_p \int_q \frac{1}{\nu_{p+q}^2}\frac{1}{\nu_p} 
\exp\left\{iq(x-y)\right\}
\nonumber\\
&&\left[-\widetilde{V}^{''}(\phi_c(x))
+\left\{-\frac{1}{2}K^{''}_0 (\phi_c(x))
(\partial_0\phi_c(x))^2
+K_0^{'}(\phi_c(x))
\partial_0\phi_c(x)ip_0
\right.\right.\nonumber\\
&&\left.-\left.\tilde K_0(\phi_c(x))(p+q)_0^2\right\}
+\left\{K_0\leftrightarrow K_s, \partial_0\leftrightarrow \nabla,
p_0 \leftrightarrow {\bf p}, q_0 \leftrightarrow {\bf q}\right\}
\right]
\nonumber\\
&\times&\left[-\widetilde{V}^{''}(\phi_c(y))
+\left\{-\frac{1}{2}K_0^{''}(\phi_c(y))
(\partial_0\phi_c(y))^2
+K_0^{'}(\phi_c(y))
\partial_0\phi_c(y)i(p+q)_0
\right.\right.\nonumber\\
&&\left.-\left.\tilde K_0(\phi_c(y))p_0^2\right\}
+\left\{K_0\leftrightarrow K_s, \partial_0\leftrightarrow \nabla,
p_0 \leftrightarrow {\bf p}, q_0 \leftrightarrow {\bf q}\right\}
\right].
\nonumber\\
&=&\int d^4x \int d^4y \int_p \int_q \frac{\exp\left\{iq(x-y)\right\}}
{\nu_{p}^3}
\label{expansion1}
\\
&&
\left\{
 1-\frac{ 
 \left\{
  \overline{K}_0 (4 p_0 q_0 + 2 q_0^2)+ 
  \overline{K}_s (4 {\bf p q} + 2 {\bf q^2}
 \right\}}{\nu_p}
 +\frac{12
 \left\{
  \overline{K}_0 p_0 q_0 + \overline{K}_s {\bf p q}
 \right\}^2}{\nu_p^2} +
 \cdots
\right\}
\nonumber\\
&&
\left[
 -\widetilde{V}^{''}(\phi_c(x))
 +\left\{
  -\frac{1}{2}K^{''}_0 (\phi_c(x)) (\partial_0\phi_c(x))^2
  +K_0^{'}(\phi_c(x)) \partial_0\phi_c(x)ip_0
 \right.
\right.
\nonumber\\
&&
\left.
 -\left.
   \tilde K_0(\phi_c(x))(p+q)_0^2
 \right\}
+\left\{
  K_0\leftrightarrow K_s, \partial_0\leftrightarrow \nabla,
  p_0 \leftrightarrow {\bf p}, q_0 \leftrightarrow {\bf q}
 \right\}
\right]
\nonumber\\
&\times&
\left[
 -\widetilde{V}^{''}(\phi_c(y))
+\left\{
  -\frac{1}{2}K_0^{''}(\phi_c(y)) (\partial_0\phi_c(y))^2
  +K_0^{'}(\phi_c(y)) \partial_0\phi_c(y)i(p+q)_0
 \right.
\right.
\nonumber\\
&&
\left.
-\left.
  \tilde K_0(\phi_c(y))p_0^2
 \right\}
+\left\{
   K_0\leftrightarrow K_s, \partial_0\leftrightarrow \nabla,
   p_0 \leftrightarrow {\bf p}, q_0 \leftrightarrow {\bf q}
 \right\}
\right],
\nonumber
\end{eqnarray}
where we take the terms up to $q^2$ since we will replace $q$ to the
derivative and need terms with second derivatives.  Similarly, by the
following exchange of the variables,
\begin{eqnarray}
\left\{
\begin{array}{lll}
p &\longrightarrow& p
\nonumber\\
q &\longrightarrow& p+q
\end{array}
\right.,
\nonumber
\end{eqnarray}
we have,
\begin{eqnarray}
&=&\int d^4x \int d^4y \int_p \int_q \frac{1}{\nu_{p+q}}\frac{1}{\nu_p^2} 
\exp\left\{iq(x-y)\right\}
\nonumber\\
&&\left[
 -\widetilde{V}^{''}(\phi_c(x))
 +\left\{-\frac{1}{2}K^{''}_0 (\phi_c(x)) (\partial_0\phi_c(x))^2
  +K_0^{'}(\phi_c(x)) \partial_0\phi_c(x)ip_0
 \right.
\right.
\nonumber\\
&&
\left.
-\left.
  \tilde K_0(\phi_c(x))(p+q)_0^2
 \right\}
+\left\{
  K_0\leftrightarrow K_s, \partial_0\leftrightarrow \nabla,
  p_0 \leftrightarrow {\bf p}, q_0 \leftrightarrow {\bf q}
 \right\}
\right]
\nonumber\\
&\times&
\left[
 -\widetilde{V}^{''}(\phi_c(y))
+\left\{
  -\frac{1}{2}K_0^{''}(\phi_c(y)) (\partial_0\phi_c(y))^2
  +K_0^{'}(\phi_c(y)) \partial_0\phi_c(y)i(p+q)_0
 \right.
\right.
\nonumber\\
&&
\left.
-\left.
  \tilde K_0(\phi_c(y))p_0^2
 \right\}
+\left\{
  K_0\leftrightarrow K_s, \partial_0\leftrightarrow \nabla,
  p_0 \leftrightarrow {\bf p}, q_0 \leftrightarrow {\bf q}
 \right\}
\right].
\nonumber\\
&=&\int d^4x \int d^4y \int_p \int_q \frac{\exp\left\{iq(x-y)\right\}}
{\nu_{p}^3}
\label{expansion2}\\
&&
\left\{
 1-\frac{
  \left\{
   \overline{K}_0 (2 p_0 q_0 + q_0^2) +
   \overline{K}_s (2 {\bf p q} + {\bf q^2})
  \right\}
 }{\nu_p}
 +\frac{4
  \left\{
   \overline{K}_0 p_0 q_0 + \overline{K}_s {\bf p q}
  \right\}^2
 }{\nu_p^2} +
 \cdots 
\right\}
\nonumber\\
&&\left[
 -\widetilde{V}^{''}(\phi_c(x))
+\left\{-\frac{1}{2}K^{''}_0 (\phi_c(x)) (\partial_0\phi_c(x))^2
  +K_0^{'}(\phi_c(x)) \partial_0\phi_c(x)ip_0
 \right.
\right.
\nonumber\\
&&\left.-\left.\tilde K_0(\phi_c(x))(p+q)_0^2\right\}
+\left\{K_0\leftrightarrow K_s, \partial_0\leftrightarrow \nabla,
p_0 \leftrightarrow {\bf p}, q_0 \leftrightarrow {\bf q}\right\}
\right]
\nonumber\\
&\times&\left[-\widetilde{V}^{''}(\phi_c(y))
+\left\{-\frac{1}{2}K_0^{''}(\phi_c(y))
(\partial_0\phi_c(y))^2
+K_0^{'}(\phi_c(y))
\partial_0\phi_c(y)i(p+q)_0
\right.\right.\nonumber\\
&&\left.-\left.\tilde K_0(\phi_c(y))p_0^2\right\}
+\left\{K_0\leftrightarrow K_s, \partial_0\leftrightarrow \nabla,
p_0 \leftrightarrow {\bf p}, q_0 \leftrightarrow {\bf q}\right\}
\right].
\nonumber
\end{eqnarray}

By comparing Eq.(\ref{expansion1}) with Eq.(\ref{expansion2}), 
we have the following "identity",
\begin{eqnarray}
0&=&\int d^4x \int d^4y \int_p \int_q \frac{\exp\left\{iq(x-y)\right\}}
{\nu_{p}^3}
\label{identity}\\
&&\left\{-\frac{ \left\{
\overline{K}_0 (2 p_0 q_0 + q_0^2) +
\overline{K}_s (2 {\bf p q} + {\bf q^2}) \right\}
}{\nu_p}
+\frac{8 \left\{
\overline{K}_0 p_0 q_0 + \overline{K}_s {\bf p q} \right\}^2
}{\nu_p^2} +
 \cdots \right\}
\nonumber\\
&&\left[-\widetilde{V}^{''}(\phi_c(x))
+\left\{-\frac{1}{2}K^{''}_0 (\phi_c(x))
(\partial_0\phi_c(x))^2
+K_0^{'}(\phi_c(x))
\partial_0\phi_c(x)ip_0
\right.\right.\nonumber\\
&&\left.-\left.\tilde K_0(\phi_c(x))(p+q)_0^2\right\}
+\left\{K_0\leftrightarrow K_s, \partial_0\leftrightarrow \nabla,
p_0 \leftrightarrow {\bf p}, q_0 \leftrightarrow {\bf q}\right\}
\right]
\nonumber\\
&\times&\left[-\widetilde{V}^{''}(\phi_c(y))
+\left\{-\frac{1}{2}K_0^{''}(\phi_c(y))
(\partial_0\phi_c(y))^2
+K_0^{'}(\phi_c(y))
\partial_0\phi_c(y)i(p+q)_0
\right.\right.\nonumber\\
&&\left.-\left.\tilde K_0(\phi_c(y))p_0^2\right\}
+\left\{K_0\leftrightarrow K_s, \partial_0\leftrightarrow \nabla,
p_0 \leftrightarrow {\bf p}, q_0 \leftrightarrow {\bf q}\right\}
\right].
\nonumber
\end{eqnarray}
Indeed, this identity holds exactly for the coefficient functions of
$(\nabla\phi_c)^2$, though it is not the case for that of
$(\partial_0\phi_c)^2$.  Anyway, we use this identity and rewrite (or
define) the third term (\ref{third}) in the following,

\begin{eqnarray}
&&\int d^4x (A^{-1} B A^{-1} B A^{-1})_{xx}
\nonumber\\
&=&\int d^4x \int d^4y \int_p \int_q \frac{\exp\left\{iq(x-y)\right\}}
{\nu_{p}^3}
\label{expansion3}\\
&&
\left
 \{1-\frac{1}{2}\frac{
  \left\{
   \overline{K}_0 (2 p_0 q_0 + q_0^2) +
   \overline{K}_s (2 {\bf p q} + {\bf q^2}) \right\}
}{\nu_p}
+ \cdots \right\}
\nonumber\\
&&\left[-\widetilde{V}^{''}(\phi_c(x))
+\left\{-\frac{1}{2}K^{''}_0 (\phi_c(x))
(\partial_0\phi_c(x))^2
+K_0^{'}(\phi_c(x))
\partial_0\phi_c(x)ip_0
\right.\right.\nonumber\\
&&\left.-\left.\tilde K_0(\phi_c(x))(p+q)_0^2\right\}
+\left\{K_0\leftrightarrow K_s, \partial_0\leftrightarrow \nabla,
p_0 \leftrightarrow {\bf p}, q_0 \leftrightarrow {\bf q}\right\}
\right]
\nonumber\\
&\times&\left[-\widetilde{V}^{''}(\phi_c(y))
+\left\{-\frac{1}{2}K_0^{''}(\phi_c(y))
(\partial_0\phi_c(y))^2
+K_0^{'}(\phi_c(y))
\partial_0\phi_c(y)i(p+q)_0
\right.\right.\nonumber\\
&&\left.-\left.\tilde K_0(\phi_c(y))p_0^2\right\}
+\left\{K_0\leftrightarrow K_s, \partial_0\leftrightarrow \nabla,
p_0 \leftrightarrow {\bf p}, q_0 \leftrightarrow {\bf q}\right\}
\right].
\nonumber
\end{eqnarray}
Note that there is no contribution from Eq.(\ref{expansion3})
to the evolution equation for the effective potential $V$
since the constant terms always include $\tilde K$ and $\tilde V''$
which are zero at $\phi_c(x)=\bar\phi$ by definition.

From now on we evaluate the coefficient functions
of $(\partial \phi_c(x))^2$.
First we calculate the first term of Eq.(\ref{expansion3}).
There are three contributions   from this term,
\begin{eqnarray}
&&\int d^4x \int d^4y \int_p \int_q \frac{\exp\left\{iq(x-y)\right\}}
{\nu_{p}^3}
\nonumber\\
&&
\left\{
 K_0^{'}(\phi_c(x))
 \partial_0\phi_c(x)ip_0 +
 K_s^{'}(\phi_c(x))
 \nabla\phi_c(x)i{\bf p}
\right\}
\nonumber\\
&\times&
\left\{
 K_0^{'}(\phi_c(y))
 \partial_0\phi_c(y)ip_0
 +K_s^{'}(\phi_c(y))
 \nabla\phi_c(y)i{\bf p}
\right\}
\nonumber\\
&&\hbox{(integrating over $q$ and $y$.)}
\nonumber\\
&=&
\int d^4x \int_p \frac{-1}{\nu_{p}^3}
\left\{
 \overline{K}_0' (\partial_0\phi_c) p_0 +
 \overline{K}_s' (\nabla \phi_c) {\bf p} 
\right\}^2,
\label{contribution11}
\end{eqnarray}

\begin{eqnarray}
&&\int d^4x \int d^4y \int_p \int_q \frac{\exp\left\{iq(x-y)\right\}}
{\nu_{p}^3}
\nonumber\\
&&
\left[
 \left\{
  K_0^{'}(\phi_c(x))
  \partial_0\phi_c(x)iq_0 +
  K_s^{'}(\phi_c(x))
  \nabla\phi_c(x)i{\bf q}
 \right\}\times\right.\nonumber\\
&&\hspace*{1cm}\left.
 \left\{
 -\tilde K_0(\phi_c(y)) p_0^2-
 \tilde K_s(\phi_c(y)) {\bf p}^2-V''(\phi_c(y))
 \right\}
\right.
\nonumber\\
&&+
\left.
 2\left\{
  K_0^{'}(\phi_c(x))
  \partial_0\phi_c(x)ip_0 +
  K_s^{'}(\phi_c(x))
  \nabla\phi_c(x)i{\bf p}
 \right\}\times\right.\nonumber\\
&&\hspace*{1cm}\left.
 \left\{
 -\tilde K_0(\phi_c(y)) p_0 q_0-
 \tilde K_s(\phi_c(y)) {\bf p q}
 \right\}
\right]
\nonumber\\
&=&
\int d^4x \int d^4y \int_p \int_q \frac{1}
{\nu_{p}^3}
\nonumber\\
&&\left[
 \left\{
  K_0^{'}(\phi_c(x))
  \partial_0\phi_c(x)\partial_{x_0} +
  K_s^{'}(\phi_c(x))
  \nabla\phi_c(x)\nabla_x
 \right\}\times\right.\nonumber\\
&&\hspace*{1cm}\left.
 \left\{
 -\tilde K_0(\phi_c(y)) p_0^2-
 \tilde K_s(\phi_c(y)) {\bf p}^2-V''(\phi_c(y)
 \right\}
\right.
\nonumber\\
&&+
\left.
 2\left\{
  K_0^{'}(\phi_c(x))
  \partial_0\phi_c(x)ip_0 +
  K_s^{'}(\phi_c(x))
  \nabla\phi_c(x)i{\bf p}
 \right\}\times\right.\nonumber\\
&&\hspace*{1cm}\left.
 \left\{
 \tilde K_0(\phi_c(y)) i p_0 \partial_{x_0}+
 \tilde K_s(\phi_c(y)) i{\bf p } \nabla_x
 \right\}
\right]
\nonumber\\
&\times&
\exp\left\{iq(x-y)\right\}
\nonumber\\
&&\hbox{(partially integrating over x and then
integrating over $q$ and $y$ .)}
\nonumber\\
&=& 
\int d^4x \int_p \frac{1}{\nu_{p}^3}
\left[
 \left\{
  \overline K_0' (\partial_0\phi_c)^2  +
  \overline K_s' (\nabla \phi_c)^2  
 \right\}
 \left\{
  \overline K_0'p_0^2+\overline K_s' {\bf p}^2 +\overline V'''
 \right\}
\right.
\label{contribution12}\\
&&\left. +2
 \left\{
  \overline K_0' (\partial_0 \phi_c) p_0 +
  \overline K_s' (\nabla \phi_c) {\bf p}
 \right\}^2
\right],
\nonumber
\end{eqnarray}
and
\begin{eqnarray}
&&\int d^4x \int d^4y \int_p \int_q \frac{\exp\left\{iq(x-y)\right\}}
{\nu_{p}^3}
\nonumber\\
&&
\left[
 \left\{
 -\tilde K_0(\phi_c(x)) q_0^2-
 \tilde K_s(\phi_c(x)) {\bf q}^2
 \right\}\times\right.\nonumber\\
&&\hspace*{1cm}\left.
 \left\{
 -\tilde K_0(\phi_c(y)) p_0^2-
 \tilde K_s(\phi_c(y)) {\bf p}^2 -V''(\phi_c(y))
 \right\}
\right]
\nonumber\\
&=& 
\int d^4x \int_p \frac{1}{\nu_{p}^3}
\left\{
 \overline K_0' (\partial_0\phi_c)^2  +
 \overline K_s' (\nabla \phi_c)^2  
\right\}
\left\{
 \overline K_0'p_0^2+\overline K_s' {\bf p}^2 +\overline V'''
\right\}.
\label{contribution13}
\end{eqnarray}
As a whole, we have the following contribution from the first term of
Eq.(\ref{expansion3}),
\begin{eqnarray}
&&\int d^4x \int_p \frac{1}{\nu_{p}^3}
\left[ 2
 \left\{
  \overline K_0' (\partial_0\phi_c)^2  +
  \overline K_s' (\nabla \phi_c)^2  
 \right\} \nu_p'
\right.
\label{contribution1}\\
&&\left. +
 \left\{
  \overline K_0' \partial_0\phi_c p_0 +
  \overline K_s' \nabla \phi_c {\bf p}
 \right\}^2
\right].
\nonumber
\end{eqnarray}
Similarly, we have the following contribution from the second term of
Eq.(\ref{expansion3}),
\begin{eqnarray}
&&\int d^4x \int_p \frac{1}{\nu_{p}^4}
\left[
-\frac{1}{2}
 \left\{
  \overline K_0'p_0^2+\overline K_s' {\bf p}^2 +\overline V'''
 \right\}^2 +
 \left\{
  \overline K_0 (\partial_0\phi_c)^2 +
  \overline K_s (\nabla \phi_c)^2
 \right\}
\right.
\nonumber\\
&&
\left.
 -2
 \left\{
  \overline K_0' (\partial_0\phi_c)p_0  +
  \overline K_s' (\nabla \phi_c) {\bf p} 
 \right\}
 \left\{
  \overline K_0 (\partial_0\phi_c)p_0  +
  \overline K_s (\nabla \phi_c) {\bf p} 
 \right\}
 \nu'_p
\right].
\label{contribution2}
\end{eqnarray}
Combining Eq.(\ref{Axx}), Eq.(\ref{ABAxx}),
Eq.(\ref{contribution1}) and Eq.(\ref{contribution2}), 
we obtain the evolution equation,  Eq.(\ref{rhs}).

\end{document}